\documentstyle[aps,prbbib,twocolumn,epsf]{revtex}

\begin{document}
\draft
\title{Parametric electron pumping through a quantum 
dot in the Kondo regime}

\author{Baigeng Wang and Jian Wang}
\address{Department of Physics, The University of Hong Kong, 
Pokfulam Road, Hong Kong, China\\
}
\maketitle

\begin{abstract}

We report a theoretical analysis of parametric electron pump  
through a quantum dot in the Kondo regime. In the adiabatic regime,
we have derived the expression for pumped current in the Kondo
regime using non-equilibrium Green's function. The pumped current 
versus different system parameters such as gate voltage, pumping
amplitude, as well as the phase difference between two pumping forces
are calculated and interesting physics are revealed.
\end{abstract}



The general physics of parametric electron pump has been the subject of 
recent studies\cite{thouless,brouwer,switkes,zhou,aleiner1,wei1,vavilov,brouwer2,sharma,aleiner,wei2,shutenko,apl,wbg1,buttiker1}
It is in particular
inspired by the recent experiment of Switkes et al. In this
experiment\cite{switkes}, the pumped current through an open quantum dot 
is driven by two gates with oscillating voltages controlling 
the deformation of the {\it shape} of the dot. The pumped DC voltage 
$V_{dot}$ is measured to vary with the phase difference $\phi$ between the 
two gate voltages, and is antisymmetric about $\phi=\pi$. At low pumping 
amplitude the experimental data gave $V_{dot}\sim \sin \phi$. In
strong pumping regime, the dependence of $V_{dot}$ on $\phi$
becomes non-sinusoidal showing $V_{dot}(0) \neq 0$, 
whereas keeping $V_{dot}(\pi)\approx 0$ for all pumping strength.
Many of these experimental findings have been explained theoretically. 
However, apart from Ref.\onlinecite{sharma,aleiner}, 
to date most of the theoretical 
investigations of parametric pumping have assumed single electron 
approximation. It would be interesting to see how the strong 
electron electron interaction modifies the pumped current. 
For this purpose, we report in this paper a theoretical analysis of 
the parametric electron pump through quantum dot in the Kondo regime
using adiabatic theory. Our results indicate, in the Kondo regime, 
that the general behavior of the pumped current is similar to 
that of the conductance. Above the Kondo temperature, as one scans 
the gate voltage $v_g$ we found two peaks in the pumped current 
corresponding to the resonant tunneling peak and Coulomb charging
peak. When the temperature is below the Kondo temperature, a new peak in 
the pumped current starts to emerge at $v_g=-U/2$ in the middle of the 
resonant peak and Coulomb charging peak.  At zero 
temperature, the pumped current has a broad peak at $v_g=-U/2$ 
which is the superposition of these three peaks. This is very different 
from the noninteracting case where there is only one peak in the 
pumped current.  In the Kondo regime, we found that as one varies the 
pumping amplitude, the pumped current increases
quadratically for small amplitude and then scales linearly with the
pumping amplitude. Our result also shows that the pumped current 
is antisymmetric about $\phi=\pi$ and is a nonsinusoidal function of 
$\phi$ for large pumping amplitude. Our result suggests that the
Kondo signature can also be found in the pumped current which 
can be checked experimentally.

We consider a 2D quantum dot with leads connected to the dot through
narrow constrictions controlled by gate voltages. Since the threshold of 
electron propagation in the constriction maybe lower than that in the
lead, the constrictions act like a double barrier whose height can be 
tuned by two gate voltages. The cyclic variation of these two
pumping gate voltages allow the parametric electron pumping through 
the quantum dot. To simply the calculation, we use the one dimensional 
double barrier potential to model the quantum dot. 

To analyze parametric quantum pumping, we make use of the nonequilibrium
Green's function method. Using the distribution function, 
the total charge in the system during the pumping is given by
$Q(x,t)=-ie\int (dE/\pi)(G^<(E,\{X(t)\}))_{xx}$
where 
$G^<$ is the lesser Green's function in real space, $x$ labels the 
position, and $\{X(t)\}$ describes a set of external parameters which 
facilitates the pumping process. $G^<$ is related to the retarded and 
advanced Green's functions $G^r$ and $G^a$,\cite{wingreen}
$~ G^<(E,\{X\}) = - f(E) [G^r(E,\{X\})-G^a(E,\{X\})]$
where the retarded Green's function in real space is 
given by

\begin{equation}
G^r(E,\{X\}) = \frac{1}{E-H-V_p -\Sigma^r}
\label{gr}
\end{equation}
In Eq.(\ref{gr}), $\Sigma^r$ is the self-energy and $V_p$ is a diagonal 
matrix describing the variation of 
the potential landscape due to the external pumping parameter $X$. 
In order for a parametric electron pump to function, we need 
simultaneous variation of two system parameters $X_1(t) =X_{10}+ X_{1p} 
\sin(\omega t)$ and $X_2(t) = X_{20}+ X_{2p} \sin(\omega t +\phi)$. 
Hence, in our case, the potential due to the gates can be written as 
$V_p=X_1 {\bf \Delta}_1+X_2 {\bf \Delta}_2$, where ${\bf \Delta}_i$ is 
potential profile for each gate. If the time variation of 
these parameters are slow, {\it i.e.} for $X(t)=X_0+\delta X 
\sin(\omega t)$, then the charge of the system coming from all contacts 
due to the infinitesimal change of the system parameter 
($\delta X \rightarrow 0$) is 

\begin{equation}
dQ(t) = \sum_i \partial_{X_i} {\rm Tr}[Q(x,t)] ~ \delta X_i(t)
\label{dq}
\end{equation}
where ${\rm Tr}[..]$ is over the positions.
It is easily seen that the total charge in the system in a period is
zero which is required for the charge conservation. To calculate the
pumped current, we have to find the charge $dQ_\alpha$ passing through 
contact $\alpha$ due to the change of the system parameters. Using 
the Dyson equation $\partial_{X_i} G^r = G^r {\bf \Delta}_i G^r$, 
Eq.(\ref{dq}) becomes,
\begin{eqnarray}
&& dQ(t) = \frac{ie}{\pi}\sum_j \int dE {\rm Tr}[G^r {\bf \Delta}_j 
G^r -c.c] f(E) \delta X_j(t)
\nonumber \\
&=&\frac{e}{\pi}\int dE (\partial_E f) \sum_j {\rm Tr}
[G^r \Gamma G^a {\bf \Delta}_j] \delta X_j(t)
\end{eqnarray}
where we have used the fact that $G^r \Gamma G^a=i(G^r-G^a)$ 
and $\Gamma=\sum_\alpha \Gamma_\alpha$ is the line width function.
So we obtain 

\begin{equation}
dQ_\alpha(t)=\frac{e}{\pi}\int dE (\partial_E f) \sum_j 
{\rm Tr}[G^r \Gamma_\alpha G^a {\bf \Delta}_j] \delta X_j(t)
\label{eq6}
\end{equation}

Furthermore, the current flowing through contact $\alpha$ due to the
variation of parameters $X_1$ and $X_2$, in one period of time, is 
given by
\begin{equation}
I_\alpha = \frac{1}{\tau} \int_0^{\tau} dt ~ dQ_\alpha/dt
\label{current}
\end{equation}
where $\tau=2\pi/\omega$ is the period of cyclic variation. In terms
of injectivity\cite{but5} given by\cite{wbg2}
\begin{equation}
\frac{dN_\alpha}{dX_j} = \int \frac{dE}{2\pi} (\partial_E f) 
{\rm Tr}[G^r \Gamma_\alpha G^a {\bf \Delta}_j] 
\label{dndx}
\end{equation} 
Eq.(\ref{current}) reduces to the familiar formula\cite{brouwer}

\begin{equation}
I_\alpha = \frac{e\omega}{\pi} \int_0^{\tau} dt \left[\frac{dN_\alpha}
{dX_1} \frac{dX_1}{dt} + \frac{dN_\alpha}{dX_2} \frac{dX_2}{dt}\right]
\label{pump}
\end{equation}
Note that Eq.(\ref{pump}) is a general expression applicable to the
case of interacting and noninteracting systems as long as the
retarded Green's function is known.

For the transport in the Kondo regime, we consider the following 
Hamiltonian $H = H_0+H_I+H_T$ with,
\begin{equation}
H_0 = \sum_{k\alpha \sigma} \epsilon_{k\alpha} C^+_{k\alpha \sigma}
C_{k\alpha \sigma} + \sum_{\sigma m} [E_m + v_g  
] d^\dagger_{\sigma m} d_{\sigma m}
\end{equation}

\begin{equation}
H_I = U \sum_m n_{m\uparrow} n_{m\downarrow}
\end{equation}
and 
\begin{equation}
H_T = \sum_{k\sigma m\alpha} T_{k\alpha m}
C^{\dagger}_{k\alpha\sigma} d_{\sigma m}
\end{equation}
where $C^\dagger_{k\alpha \sigma}$ is the creation operator of lead
$\alpha$ and $d^\dagger_{\sigma m}$ is the creation operator of the
scattering regime at energy level $m$. We have applied the gate voltage 
$v_g$ to control the energy level in the scattering region.
For this Hamiltonian $\Gamma_\alpha$ defined in Eq.(\ref{eq6}) is given
by $(\Gamma_\alpha)_{mn}=2\pi\sum_k T^*_{k\alpha m} T_{k\alpha n}
\delta(E-\epsilon_{k\alpha})$. 
There are many approaches to treat scattering problem in Kondo 
regime\cite{yeyati,kotliar,lee,langreth,nca}. 
We find it is convenience to use the perturbation
scheme proposed by Levy Yeyati et al\cite{yeyati} and Kajueter and 
Kotliar\cite{kotliar}. In this approach, the retarded Green's function 
is given by

\begin{equation}
G^r = \frac{1}{E-H -\Sigma^r_{lead}-\Sigma^r_s}
\label{gr1}
\end{equation}
where $\Sigma^r_{lead}$ is the self-energy due to the coupling between the 
scattering region and leads.  The effect of strongly electron-electron 
interaction is included in the self-energy 
$\Sigma^r_s$\cite{yeyati,kotliar,foot1} 

\begin{equation}
\Sigma^r_s(E) = Un + \frac{A\Sigma^r_0(E)}{1-B \Sigma^r_0(E)}
\label{sigma}
\end{equation}
where $\Sigma^r_0$ is the self-energy due to the second order contribution
in $U$,

\begin{eqnarray}
&& \Sigma^r_0(E) = \frac{i U^2}{8\pi^3} \int \frac{dE_1 dE_2 dE_3}
{E+E_3-E_1-E_2+i\delta} \times \nonumber \\
&& [G_0^>(E_1) G_0^>(E_2) G_0^<(E_3) - G_0^<(E_1) G_0^<(E_2)
G_0^>(E_3)]
\end{eqnarray}
where $G_0^r=1/(E-H_0-\Sigma^r_{lead})$ and $G_0^<=-f(G_0^r-G_0^a)$.
Here for simplicity, we have only considered a particular energy level 
$E_0$ and used the wideband limit\cite{wingreen}.
The coefficients $A$ and $B$ in Eq.(\ref{sigma}) are determined by the
solutions in two limiting cases: large energy limit and atomic 
limit\cite{kotliar}, from which we have 
$A=[n(1-n)]/[n_0(1-n_0)]$
and 
$B = [(1-2n)]/[n_0(1-n_0) U]$
with $n = -\int dE f(E) {\rm Im} G^r/\pi$ is the physical particle
number and $n_0 = -\int dE f(E) {\rm Im} G^r_0/\pi$ is the fictitious 
particle number. This scheme gives a good description for the case of 
half filling. Away from that, one must replace $H_0$ in $G^r_0$ and
$G^r$ by a self-consistent Hamiltonian $H_{eff}$ and use the Friedel 
sum rule\cite{yeyati,kotliar}

\begin{equation}
n = \frac{1}{2} - \frac{1}{\pi} \arctan[\frac{E+\Sigma^r_s + 
{\rm Re}\Sigma^r_{lead}} {{\rm Im}\Sigma^r}]
\label{sum}
\end{equation}
The self-consistent solution of Eqs.(\ref{gr1}), (\ref{sigma}), and 
(\ref{sum}) determines the self-energy $\Sigma^r_s$ which will be used
in the calculation of pumped current.  We now apply Eq.(\ref{current}) 
to calculate the pumped current in the Kondo regime.
The double barrier structure is modeled by potential 
$U(x)=X_1 \delta (x+a/2)+X_2 \delta (x-a/2)$ where $X_1$ and $X_2$ are 
barrier heights which varys in a cyclic fashion to allow the charge
pumping. In particular, we set $X_i = v_0+v_p\sin(\omega t+\phi_i)$
with $\phi_1=0$ and $\phi_2=\phi$. We will fix the units
by setting $\hbar=2m=1$ in the following analysis. For the GaAs system 
with $a=1000A$, the energy uint is $E=56 \mu eV$. 
We will also fix the on site potential $U=5$ which is much smaller
than the level spacing in the quantum dot, frequency $\omega=1$,
the barrier height $v_0=79.2$, and phase difference $\phi=\pi/2$  
(unless specified otherwise). Finally, the energy of incoming
electron is chosen to be in line with a resonant level $E_0$ when
$v_g=0$. In Fig.1 we present the transmission 
coefficient versus gate voltage (which controls the levels in the 
quantum dot) at different temperatures $T$. 
When the temperature is higher than the Kondo temperature $T_k=0.02$
(dashed line in Fig.1), we see two peaks: resonant tunneling peak at 
$v_g=0$ (for $E=E_0$) and the Coulomb charging peak at $v_g=-U$
($E=E_0+U$). At low temperatures below $T_k$, the co-tunneling process 
leads to a new peak, the Kondo peak, at the Fermi level. As the
temperature is lowered, the peak height of Kondo peak increases
and the dip between resonant peak and charging peak diminishes. 
At zero temperature, the broad peak at $v_g=-U/2$
in Fig.1 is the superposition of these three peaks.  For $v_g>0$ or 
$v_g<-U$, the transmission coefficient is almost temperature independent. 
Note that the peak heights 
(near $v_g=0$ and $v_g=-U$) are asymmetric about the 
$v_g=-U/2$. This is because the linewidth function $\Gamma$ depends 
on energy or in our case depends on the gate voltage. Fig.2 depicts the 
pumped current as a function of gate voltage at different temperatures and
for different pumping amplitudes. Generally speaking, the pumped current
follows similar pattern of the transmission coefficient at different 
temperatures due to the fact that the pumped current is proportional to
the density of states of the system which also manifests in the 
transmission coefficient. We see that as the pumping amplitude becomes
larger, the ratio $I_p(T=0,v_g=-U/2)/I_p(T=1.5T_k,v_g=-U/2)$ becomes smaller
(see Fig.3 for further discussion);  
at $T \neq 0$ two resonant peaks become broader and move away from
each other; the Kondo peak at
$T=0$ becomes broader and flattened. We also notice that 
the pumped current increases as the pumping amplitude increases. 
In particular, as the pumping amplitude increases, the peak height
of resonant states at $E_0$ and $E_0+U$ increase much faster than 
that of the Kondo peak and when $v_p=0.1 v_0$ they have almost 
the same height.
In Fig.3, we plot the pumped current versus relative pumping amplitude
$v_p/v_0$ at two different gate voltage: one at 
$v_g=-U/2$ and the other near the resonant level when $v_g=-0.5$. 
At the $v_g=-U/2$, the dependence on the
relative pumping amplitude shows the expected quadratic behavior for
small amplitude since the pumped current is bilinear in pumping
amplitude in the weak pumping regime\cite{brouwer}. For larger 
amplitude $v_p/v_0>0.03$ it is almost linear with different 
slopes depending on temperatures. The slope is smaller at higher 
temperature. For the gate voltage near the resonant level, the
pumped current has similar behavior except that it is not sensitive
to the change of temperature. 
Fig.4 displays the pumped current as a function of phase difference
$\phi$ between two pumping forces for different pumping
amplitudes. The pumped current is antisymmetric about the phase
difference $\phi=\pi$. In the weak pumping regime ($v_p=0.01v_0$), the 
pumped current shows the sinusoidal behavior and peaked at $\phi=\pi/2$. 
This is because in the weak pumping regime, the pumped current is
bilinear in the pumping amplitude and proportional to 
$\sin\phi$.\cite{brouwer}  In the strong pumping regime ($v_p=0.1v_0$), 
we start to see nonsinusoidal behavior as higher order terms of pumping
amplitude come into play. The maximum pumped current occurs
approximately at $\phi=0.6\pi$ (see Fig.4a and Fig.4b).  
Similar nonlinear
behavior is also seen experimentally\cite{switkes} although 
the physical origin may be different. 

In summary, we have studied the parametric electron pumping through
a quantum dot in the Kondo regime using a nonequilibrium Green's 
function theory. We found that the behavior of the pumped current is
closely related to the conductance. As one varies the pumping amplitude, 
the pumped current increases quadratically for small amplitude and 
then scales linearly with the pumping amplitude.
Because of the resonant nature of the pumping, the pumped current shows
nonsinusoidal dependence on the phase difference of the pumping
parameters.
In this paper, we have used the adiabatic theory to calculate the
pumped current. This theory is valid in the low frequency regime and
can not account for the anomaly at $\phi=\pi$ found 
experimentally\cite{switkes}. At finite frequency, one must use
the real space nonequilibrium Green's function method\cite{wbg1} 
to calculate the pumped current.

\section*{Acknowledgments}

We gratefully acknowledge the support by RGC of Hong Kong SAR under
grant number HKU 7091/01P. 



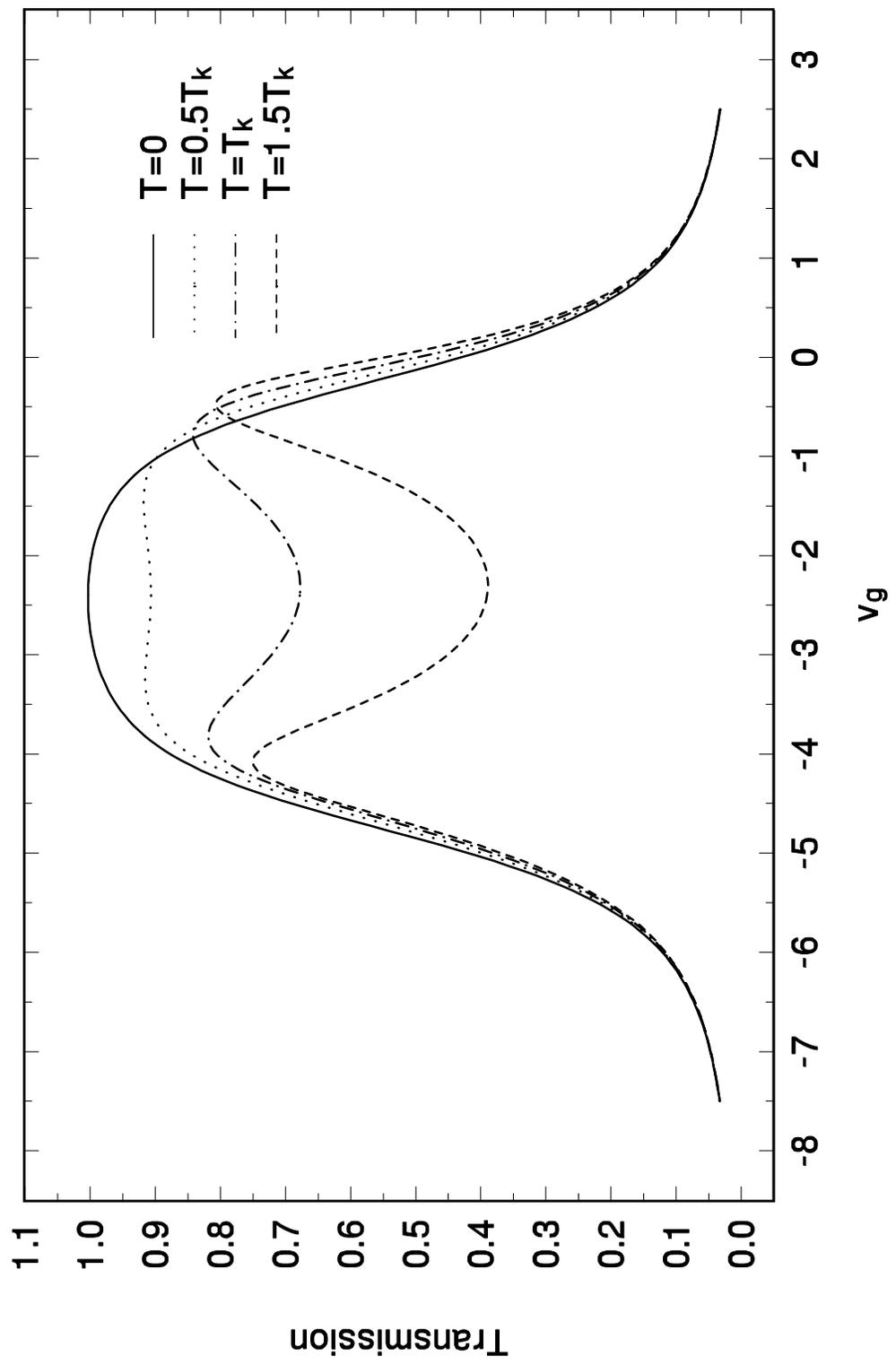
\begin{figure}
\caption{The transmission coefficient versus gate voltage at different
temperatures. 
}
\end{figure}

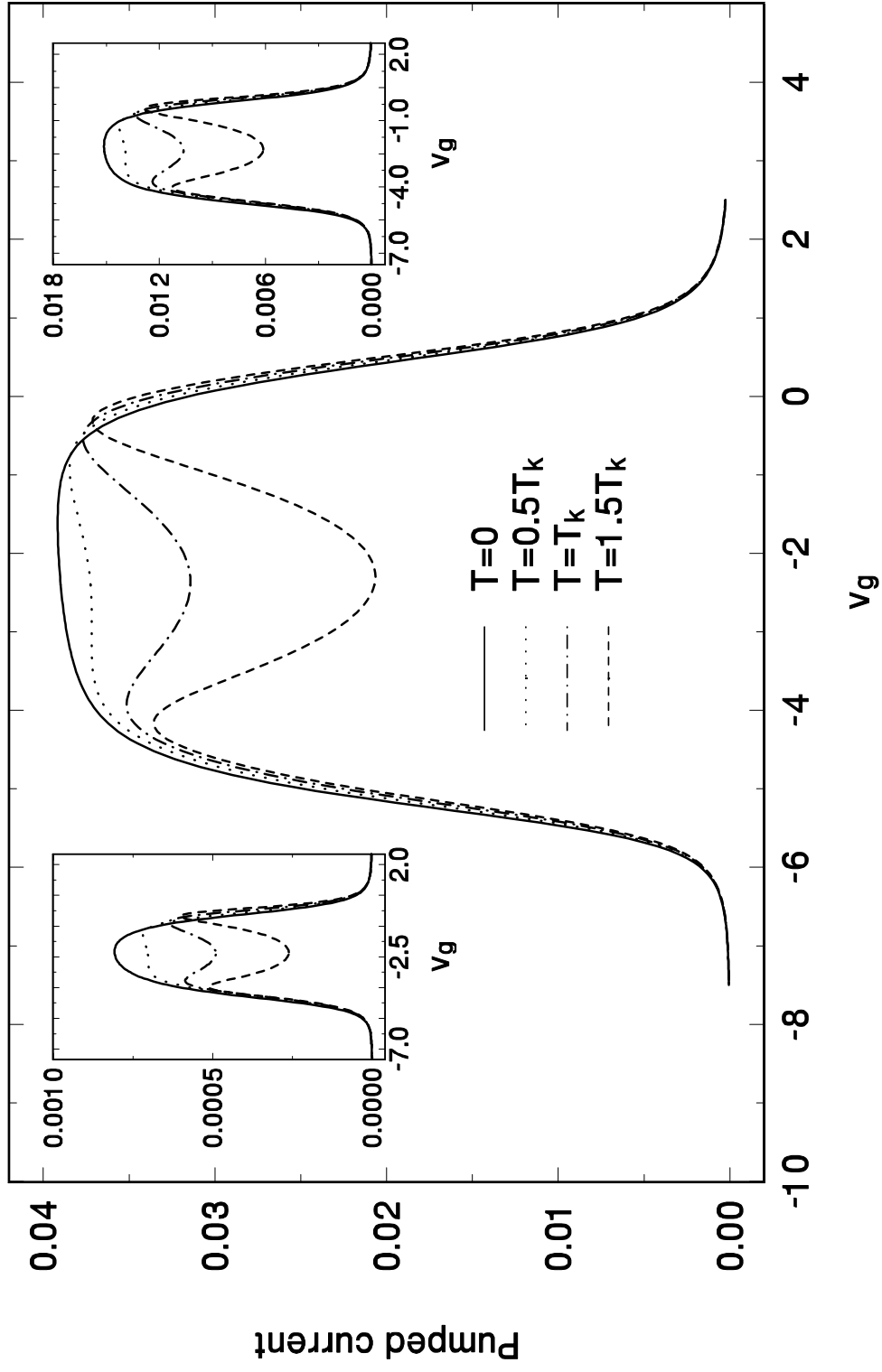
\begin{figure}
\caption{The pumped current versus gate voltage for different pumping
amplitudes $v_p$. Main figure: $v_p=0.1 v_0$; left inset: 
$v_p=0.01 v_0$; right inset: $v_p=0.05 v_0$.
}
\end{figure}

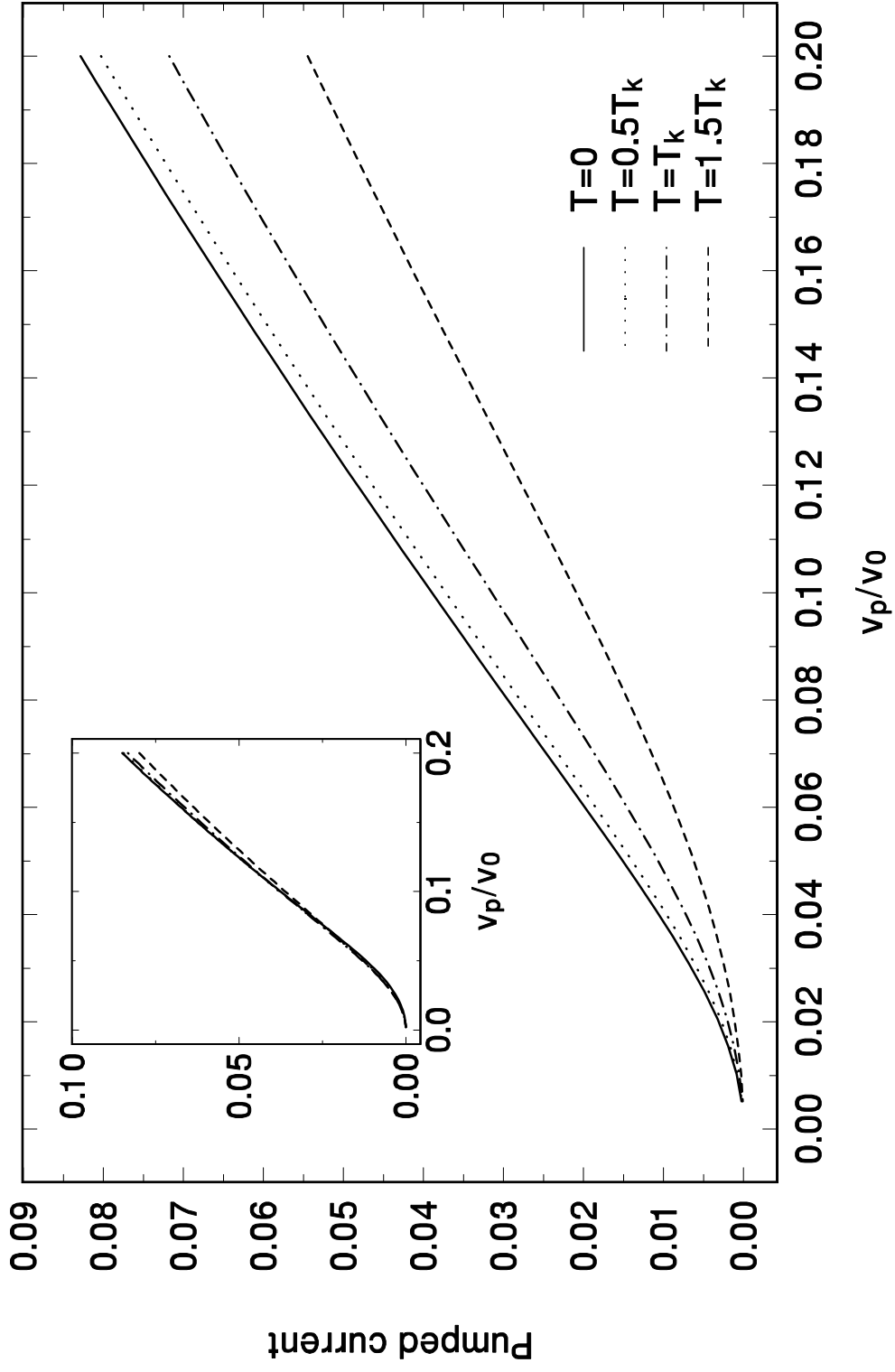
\begin{figure}
\caption{The pumped current versus relative pumping amplitude at different 
gate voltages.  Main figure: $v_g=-2.5$; inset: $v_g=-0.48$.
}
\end{figure}

\begin{figure}
\caption{The pumped current versus phase difference at different 
temperatures.  (a). Main figure: $v_g=-2.5$ and $v_p=0.01 v_0$; 
inset: $v_g=-2.5$ and $v_p=0.1 v_0$. (b). Main figure: $v_g=-0.48$ 
and $v_p=0.01 v_0$; inset: $v_g=-0.48$ and $v_p=0.1 v_0$.
}
\end{figure}



\begin{thebibliography}{00}
\bibitem{foot1}
Similar to Ref.\onlinecite{kotliar}, we have assumed that there
is no magnetic symmetry breaking ($n_\sigma=n_{{\bar\sigma}}=n$).
\end{thebibliography}
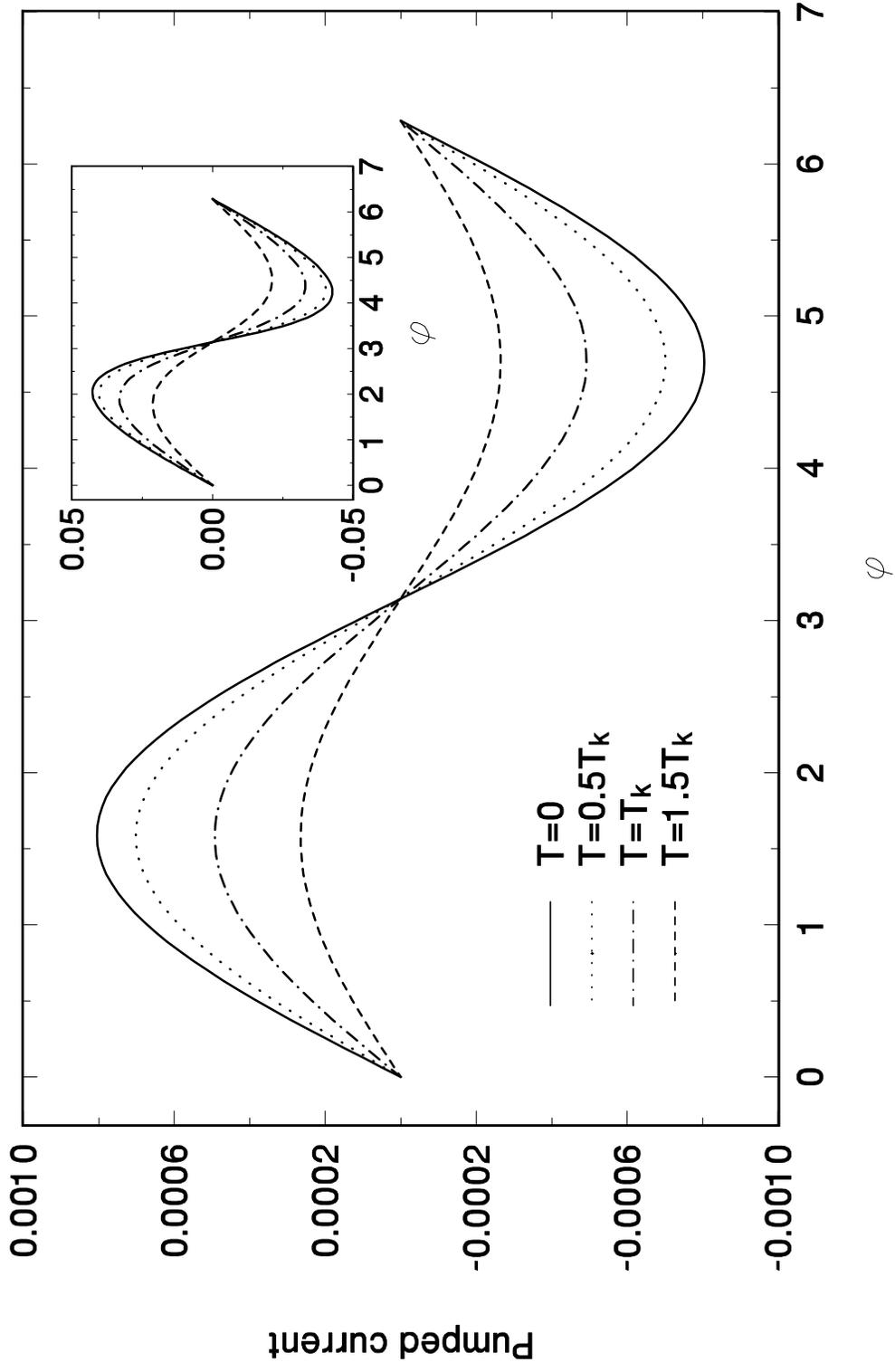
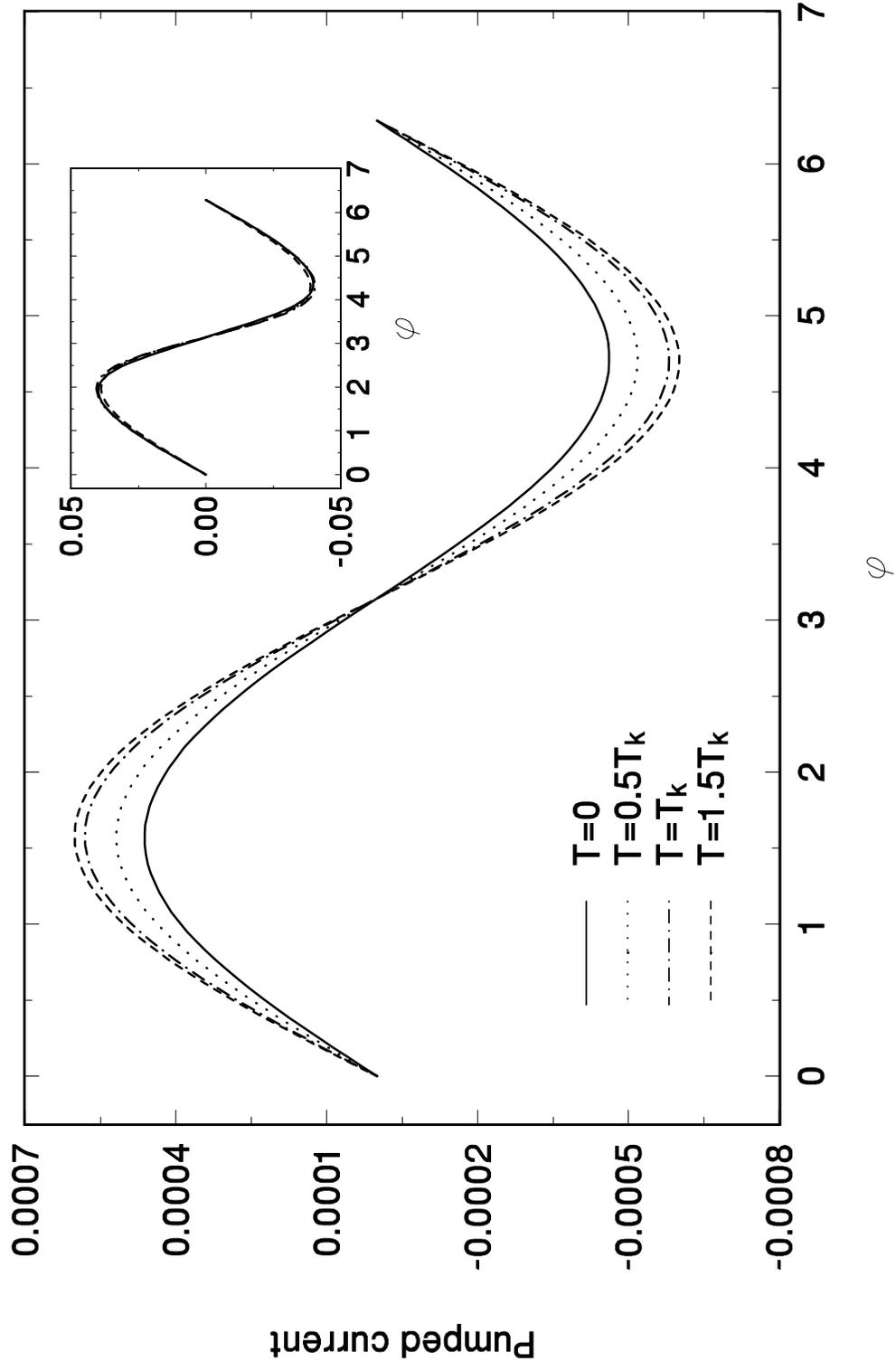
\end{document}